# A Compact Dispersive Refocusing Rowland Circle X-ray Emission Spectrometer for Laboratory, Synchrotron, and XFEL Applications

William M. Holden[1], Oliver R. Hoidn[1], Alexander S. Ditter[1], Gerald T. Seidler[1,*], Joshua Kas[1], Jennifer L. Stein[2], Brandi M. Cossairt[2], Stosh A. Kozimor[3], Jinghua Guo[4], Yifan Ye[4], Matthew A. Marcus[4], Sirine Fakra[4].

[1]Physics Department, University of Washington, Seattle WA

[2]Chemistry Department, University of Washington, Seattle WA

[3]Los Alamos National Laboratories, Los Alamos New Mexico

[4]Advanced Light Source, Lawrence Berkeley National Lab, Berkeley CA

X-ray emission spectroscopy is emerging as an important complement to x-ray absorption fine structure spectroscopy, providing a characterization of the occupied electronic density of states local to the species of interest. Here, we present details of the design and performance of a compact x-ray emission spectrometer that uses a dispersive refocusing Rowland (DRR) circle geometry to achieve excellent performance for the 2 – 2.5 keV range, i.e., especially for the *K*-edge emission from sulfur and phosphorous. The DRR approach allows high energy resolution even for unfocused x-ray sources. This property enables high count rates in laboratory studies, comparable to those of insertion-device beamlines at third-generation synchrotrons, despite use of only a low-powered, conventional x-ray tube. The spectrometer, whose overall scale is set by use of a 10-cm diameter Rowland circle and a new small-pixel CMOS x-ray camera, is easily portable to synchrotron or x-ray free electron beamlines. Photometrics from measurements at the Advanced Light Source show somewhat higher overall instrumental efficiency than prior systems based on less tightly curved analyzer optics. In addition, the compact size of this instrument lends itself to future multiplexing to gain large factors in net collection efficiency, or its implementation in controlled gas gloveboxes either in the lab or in an endstation.

(*) Corresponding author: seidler@uw.edu

# I. Introduction

High-resolution x-ray emission spectroscopy (XES) has demonstrated its utility across a wide range of contemporary problems.[1-22] However, the limitation of this method to synchrotron light sources has inhibited its broader implementation, especially for industrial and more analytical, rather than fundamental, directions. Recent work in several groups[23-28] has aimed to resolve this issue by developing laboratory-based XES instruments ranging from as low as the C K-edge (284 eV)[27] to as high as Au K-edge (78 keV).[28] Of particular interest here, several groups have made high-resolution laboratory measurements of sulfur and phosphorous x-ray emission using double-crystal spectrometers,[29-31] von Hamos geometry instruments,[32,33] and a dispersive Rowland circle geometry.[34-37] These spectrometers were all built in the latter half of the twentieth century, and required multi-kW x-ray tubes to achieve reasonable measurement times. More recent laboratory-based work in this energy range has seen an impressive extension to proton-induced x-ray emission analysis,[38, 39] with the same Rowland-circle instrument also seeing important use at a synchrotron endstation.[17, 18, 40]

Here we present technical details and representative results for an extremely compact, highly efficient, dispersive x-ray emission spectrometer designed to function particularly well in this 2-2.5 keV energy range relevant for S and P K-shell emission. We present measurements on S and P XES both in the laboratory and at a synchrotron endstation. Our results include a representative, laboratory-based analytical application where we determine the distribution of oxidation states of P in InP quantum dots. In the laboratory setting, this instrument also has the important feature that it can be operated very efficiently with an unfocused x-ray source and consequently a large beamspot on the sample. This greatly decreases cost and increases ease of use. When implemented at the synchrotron, either a focused or unfocused source can be used, with the former giving a modest improvement in energy resolution.

The performance and small size of the instrument reported here results from a synergistic overlap of important technical features of the x-ray analyzer, the position-sensitive detector, and the overall optical configuration itself. High-quality cylindrically-bent crystal analyzers are only recently commercially available with extremely small, i.e., 10-cm, radii of curvature. In general, the overall size of any Rowland-circle x-ray spectrometer scales linearly with the radius of curvature of the optic, so the decrease from the 'standard' 1-m radius of curvature spherically-bent crystal analyzer, as commonly used in synchrotron endstation instrumentations[41-44] and also used in lab-based spectrometers for the hard x-ray range,[23] to the present 10-cm radius optic allows a corresponding factor of 10 change in the linear dimensions of the resulting instrument. The compact size offers advantages in portability and ease of installation for use at synchrotron and XFEL endstations, as well as a unique scientific possibility: it is small enough that it could be readily integrated into controlled-gas glove box systems to enable new directions in analytical chemistry for air-sensitive materials.

Concerning the position-sensitive detector, our recent and ongoing development of complementary metal-oxide-semiconductor (CMOS) x-ray cameras using mass-produced sensors is a critical enabling technology.[45, 46] These sensors, including the back-illuminated sensor used in the present instrument, can have pixels below 3-μm pitch while also having useful spectroscopic capability in the 2-5 keV energy range. The small pixel size is necessary if fine energy resolution is to be maintained for the tightly curved Bragg optic, and the sensitivity to the energy of each recorded incident photon allows for very high rejection of background signals, minimizing the need for internal shielding.

Finally, as a central defining concept of the instrument design, we employ a dispersive refocusing Rowland (DRR) geometry, defined in detail in Section II below. In the context of XES, we believe that this geometry was first pointed out by Dolgih and Yarmoshenko *et al.*,[34] who constructed a DRR spectrometer based on a much larger 1.3 m Rowland circle. The DRR optical layout has considerable qualitative similarities to that for Bragg-Brentano diffractometers using position sensitive detectors.[47, 48] The DRR approach is beneficial in three practical matters: (1) its insensitivity to the illumination spot size allows the use of conventional 'x-ray fluorescence (XRF) style' x-ray tubes without any focusing optic; (2) the option of larger illuminated spot size (as much as ~5 mm) on the sample helps to reduce the exposure flux density and consequently serves to lower the likelihood of x-ray induced damage during the measurement; and, (3) the resulting spectrum, if a large beamspot is used, constitutes an average over the illuminated region, giving a decreased sensitivity to spatial inhomogeneities in sample preparation.

We continue as follows. First, in section II, we describe the general ray-tracing and design issues that arise in the DRR geometry. This includes a discussion of the similarities and differences between the DRR geometry and other more common Rowland-circle approaches. Second, in section III we give a detailed description of the implementation of the instrument in the laboratory and synchrotron environments, and provide other experimental details. Next, in section IV we present results taken in the laboratory and at the synchrotron, comparing and contrasting energy resolution and count rates in the present study with those previously reported at synchrotron light sources. This includes a representative laboratory-based 'analytical' application where we quantify the fractional oxidation of P in InP nanocrystalline materials. Finally, we conclude and discuss future directions in Section V.

# II. The Dispersive Refocusing Rowland (DRR) Geometry

To operate a dispersive spectrometer using a Rowland-circle geometry, there are three basic configurations using different source sizes and locations, as shown in Fig. 1. In Fig. 1a, where the sample is illuminated by a small point-source *off* of the Rowland circle, multiple energies will be diffracted from the crystal. These energies diffract at different angles and ray-tracing back towards the Rowland circle results in a collection of 'virtual' point sources on the source arc of the Rowland circle.[49] The bandwidth diffracted then depends on the distance of the sample from the crystal analyzer.[50] In the limit when the source is significantly off circle, the location of the position-sensitive detector (PSD) is not especially delicate – the ray-tracing is non-focal, and diffracted rays have only a very small divergence. Because of this small divergence, the PSD need not be tangent to the Rowland circle to achieve high-resolution.[49] Each energy diffracts from a very small portion of the crystal, resulting in no appreciable Johann error, even at low Bragg angles. The geometry of Fig. 1a is used in the synchrotron results of section IV.B.

In Fig. 1b, the situation is shown for large-spot illumination of a sample *on* the Rowland circle. In this case, different energies of fluoresced x-rays are collected from different portions of the sample, thus requiring a homogeneous sample to avoid gross systematic error in the dispersed spectrum on the detector arc. For each point of the sample on the source arc of the Rowland circle, x-rays of a particular energy are captured by the *entire* crystal. This increases signal strength at each energy, but at the cost of larger divergence of the rays refocusing onto the detector, as well as an increased potential for Johann error at lower Bragg angles. The large divergence of the refocused, analyzed radiation requires precise placement of the camera for a high energy-resolution signal, i.e., the depth of focus is small. With the incident beam brought in perpendicular to the Rowland plane, this geometry is used for combined imaging and spectroscopic measurements in many plasma physics studies, as well as in some synchrotron applications.[51-53]

Finally, in Fig. 1c and of direct relevance here, a large-spot illumination of a sample *off* the Rowland circle has an effective collection of 'virtual' sources at different energies, as in panel (a). However, for a large beamspot, each portion of the sample can contribute at all energies, except when truncated at the boundary of the analyzer. This 'dispersive Rowland refocusing' (DRR) approach removes the need for a focused beam, while also decreasing the sensitivity to sample inhomogeneities by giving a natural averaging of the spectrum over the illuminated region of the sample. Spectrometers making use of this approach have been recently discussed and implemented at some synchrotron endstations,[50] where the relatively large spot size allows for efficient measurements, e.g. for gaseous samples.[40, 54] In the present spectrometer, the large spot size is taken to an even greater extreme, allowing the use of an unfocused x-ray tube on an extremely small 10-cm Rowland circle.

To minimize distortions of the spectrum in the DRR geometry, it is important that each energy in the bandwidth of interest has equal net detection efficiency in the final spectrum. This can be accomplished in either of two extremes: (1) the sample illumination is extremely large such that each energy is usefully captured by the entire crystal analyzer, or (2) the sample illumination is sufficiently small that each energy uses the same fraction of the crystal, i.e. that each illuminated sub-region of the sample can contribute at all energies from some segment of the crystal analyzer. Within either configuration, distance from the crystal to the sample can be varied to trade-off count rates against total energy bandwidth analyzed.

Moving now to the specific implementation of mechanical components, there are numerous possible implementations of any Rowland-circle configuration, with the pragmatically-preferred approach always determined by external criteria. Here, we choose to fix the x-ray source in the lab frame to simplify operation in both the laboratory and the synchrotron. In addition, we choose to let the circle be free to rotate about the source axis so that the camera face can always be kept parallel to the same reference plane, thus minimizing the number of degrees of freedom that require fine-tuning. These characteristics are illustrated in Fig. 2. After coarsely adjusting the height of the camera so that the analyzed radiation strikes the sensor, necessary fine-adjustments are made to bring the sensor tangent to the Rowland circle and hence to the refocal position for the analyzed radiation.

Returning to Fig. 2, when a new Bragg angle is desired, in order to keep the source location fixed and maintain the camera orientation, the Rowland circle must be moved by appropriate modification to the location and rotation of the crystal analyzer. Finally, note that this requires that the distance of the focal point region on the Rowland circle from the spectrometer chamber wall varies for different Bragg angles. We show below, in the detailed instrument design, that this is easily addressed.

# III. Experimental

## III.A Laboratory Environment

Computer-aided design (CAD) renderings and photographs of the spectrometer, as implemented in the laboratory environment, are shown in Figures 3 and 4.The sample is directly illuminated with x-rays from a conventional, air-cooled tube source (Varian VF-50 with a Pd anode) having a maximum electron beam power of 50W at 25 kV accelerating potential. The choice of a Pd anode is advantageous due to its strong fluorescence lines at ~2.8 keV that are very effective at stimulating *K*-shell photoionization of P and S. The x-ray tube is driven by a Spellman uX50P50 high voltage power supply. The VF-50 provides an *unfocused* beam of combined bremsstrahlung and characteristic fluorescence radiation from the Pd anode.

Fluorescence from the sample is diffracted by a 10-cm radius, cylindrically-bent, Si (111) Johann analyzer (XRS Tech) in the DRR geometry of the corresponding 10-cm diameter Rowland circle. The analyzer dimensions are 20 mm (width) x 8 mm (height, out of the Rowland plane). The Si (111) orientation provides Bragg angles of 79° for P Kα (2014 eV) and 59° for S Kα (2308 eV). For P, the sample is placed in the second configuration described in section II, such that each energy in the resulting spectrum makes use of the same fraction of the crystal. When measuring S K-shell XES, however, the relatively low Bragg angles introduce a large Johann error distorting the spectrum. To compensate, the edges of the crystal analyzer are masked with aluminum foil so that only the central ~4 mm are used. This results in the first configuration described in section II, wherein the fluorescence from the sample now makes use of the entire 4 mm of the crystal for all analyzed energies. Moving to a Johansson-type analyzer would clearly improve efficiency for S XES.

Placement of the crystal analyzer is accomplished using 3-D printed plastic mounting pieces that register with the walls of the spectrometer box to determine the optics position and orientation, as per the discussion in section II. Again, the location of the crystal is chosen such that the camera, oriented vertically outside of the chamber, is tangent to the Rowland circle at the energy of interest. Also, as described in section II, alignment and tuning is achieved by first locating the signal on the camera via manual, vertical translation, and then by fine-focusing adjustments where the camera is moved closer to or further away from the spectrometer chamber.

The refocused rays are detected by a recently developed energy-resolving x-ray camera[46], which uses a back-illuminated CMOS sensor and is similar to a previously-reported instrument.[45] The detector is based on a commercial amateur astronomy camera (ZWO Company) that we have modified by removing the glass from its image sensor (Sony IMX-291), allowing x-rays to directly illuminate the sensor's active region. The camera's CMOS sensor has a pixel pitch of 2.9-μm and a 1936×1096 pixel layout. On a 10-cm radius Rowland circle, the sensor's 2.9-μm pixel size corresponds to an energy broadening of ~0.01 eV at P Kα and ~0.04 eV at S Kα. The charge separation generated by each photon absorbed on each pixel results in a proportional readout value. In the simplest case, wherein the entire charge cloud from an incident x-ray event is concentrated in a single pixel, the detecting pixel has intrinsic sensitivity to the energy of the incident x-ray photon. In the majority of events, however, the charge cloud spreads over a cluster of several adjacent pixels.[45, 46] To include all events while preserving optimal energy resolution we have developed software to identify such clusters and reconstruct corresponding photon energies and positions.[45, 46, 55] With this additional processing, the sensor's quantum efficiency (QE) and energy resolution are 65% and 150 eV, respectively, at the photon energy of P Kα. The quantum efficiency of the sensor decreases considerably above 4 keV,[46] leading us to focus here on the 2 – 2.5 keV energy range where the Si 111 optic and camera performance are synergistic.

Note that the use of an energy-resolving, single photon counting camera is extremely beneficial in the present application. Frame-by-frame rejection of single-photon events that are outside the energy window of interest removes a wide range of backgrounds that would otherwise contaminate an energy-integrating position-sensitive detector. Consequently, only very minimal internal shielding is required.

From the discussion in Section II, the fluorescence x-rays from the sample are dispersed onto the detector, and each pixel acts as an effective slit that only accepts rays in a narrow energy band. For a cylindrical optic, like that used here, the focusing is theoretically exact in the plane of the Rowland circle, but for rays with some out-of-plane divergence, the rays are bent towards the backscatter (low-energy) direction. The results in a curved focal line[40] as the ideal point-source response function. The curvature itself does significantly affect instrumental resolution since the curved focal line can be taken into account to produce a spectrum, as is commonly done in synchrotron implementations.[40, 54, 56] In the present spectrometer in the laboratory setting, however, the curvature of the signal convolved with the large spot illumination results in irreversible blurring of the ideal point-source response function for regions of the sensor far out of the Rowland plane. In order to maintain high resolution for lab-based measurements, this requires that the signal on the detector be cropped around the region centered on the Rowland plane to exclude these blurred regions.

The absorption length in air of, e.g., S Kα fluorescence, is ~1.5 cm. Consequently, the sample, crystal analyzer, and majority of the beam path are inside of a 30 × 30 × 7 $cm^3$ aluminum chamber which is operated at rough vacuum (~200 mTorr) or else flushed and filled with He at 1 atm. As shown in Figures 2, 3, and 4, the camera is outside of the spectrometer box. The x-rays exit through an 8-μm thick polyimide film and traverse a ~2-3 mm air gap before landing on the camera. The small air-gap is maintained by having different thickness spacers below the polyimide window. The combined absorption from the polyimide exit window and small air path is ~50 – 60% in the targeted 2 – 2.5 keV energy range, leaving room for a two-fold increase in counts if the camera is adapted to mount inside of the spectrometer chamber.

The energy scale of the measurements was determined by ray-tracing considerations, using the 10-cm Rowland circle geometry. Measuring samples on the same energy scale (including one or more reference materials) is made possible by swapping samples while maintaining the analyzer and camera locations. After an ensemble of internally-consistent results is obtained, a single common energy shift is applied all spectra so as to match, e.g., prior published results for one of the reference materials.[18, 57]

## III.B. Synchrotron Environment

The compact size of the spectrometer allows for easy transport and interface with existing synchrotron and x-ray free electron laser beamlines. To demonstrate this utility, the spectrometer was taken to bending-magnet beamline 10.3.2 at ALS for demonstration studies. Installation was straightforward, requiring less than two hours from the start of setup to the onset of data collection. The spectrometer fit within the existing endstation equipment, and the laboratory x-ray source was replaced by a simple adapter connected to the beampipe using flexible bellows. Helium gas was flowed through the chamber and bellows throughout the measurements. The configuration of the various spectrometer components is otherwise unchanged from the laboratory setup in section III.A, above. This allowed for the spectrometer to be pre-calibrated and pre-focused for a fluorescence line of interest (in this case P Kα) before traveling to the synchrotron.

The incident flux was $3\times10^9$/s at the selected incident photon energy of 3 keV. This was determined using a gas ionization chamber with an effective path length of 2 cm, filled with a 33% $N_2$, 67% He mix and also confirmed with a second measurement using only pure $N_2$, in the ion chamber. The spectrometer was translated downstream from the usual focus of this microfocus beamline to achieve a ~200-μm spot size on the sample. This small source size simplified treatment of the sensor image. Whereas in laboratory operation blurring of the image out of the Rowland plane requires significant cropping of the signal to achieve high-resolution, the small spot size at the beamline resulted in a signal that required minimal cropping. Despite the reduction in count rates due to the low flux of the bending magnet source, background signals were proportionally reduced and good spectra were achieved with longer integration times.

### III.C Samples

For testing the spectrometer in the laboratory and at the synchrotron, the following phosphorous- and sulfur-containing reference samples were used: a $10\times10\times0.5$ mm$^3$ crystalline wafer of GaP (MTI Corporation), a $10\times10\times1$ mm$^3$ crystalline wafer of ZnS (MTI Corporation), and a pressed powder pellet of $FePO_4$ (Alfa Aesar) mixed with graphite binder. For the pellet, $FePO_4$ powder was mixed with graphite in an approximate 1:3 ratio, then pressed into a 13 mm-diameter pellet.

P Kα emission for three different samples of InP nanocrystals was measured as a pilot study to demonstrate the analytical capabilities of the spectrometer. The InP quantum dots (QDs) were prepared by a procedure from Gary and Cossairt.[58] One sample consisted of the pure as-synthesized InP QDs, while the second sample consisted of Zn-passivated InP QDs, which were post-synthetically modified with zinc oleate as described in Stein *et al*.[59] The third sample measured was InP magic-sized clusters with the initial carboxylate ligand environment replaced with phosphonates, prepared following a procedure from Gary *et al*.[60]

## IV. Results and Discussion

To begin, a 500 s exposure for P Kα emission of GaP is shown in Fig. 5. As mentioned in section III.A., the blurring due to finite sample illumination size requires cropping the image to the central 600 rows to retain high energy resolution. We show measurements of the P Kα emission spectrum for GaP and $FePO_4$ in Fig. 6. An energy shift of 1.03 eV is observed due to the difference in oxidation state, $P^{3-}$ in GaP, and $P^{5+}$ in $FePO_4$. The integration time of the measurements was 500 s, although for these concentrated samples the shift is clearly observed after only 60 s of measurement. The energy bandwidth depends on the spectrometer geometry, and at the P Kα Bragg angle of 79° the bandwidth is only 12 eV. The GaP sample gave a useful count rate of 1200/s and the $FePO_4$ pellet, being somewhat less concentrated, gave a useful count rate of 450/s. The count rate for GaP is impressive, in that it is ~50% that observed in a prior synchrotron-based study using a third-generation insertion device beamline.[57]

P Kβ emission was measured on the same samples, and the results are shown in Fig. 7 (top). Although the peak Kβ signal is ~40× weaker than that of the Kα, clean spectra are measured in ~600 s. As the Kβ emission constitutes valence-to-core transitions for P and S, it is sensitive to the chemical bonding environment. For example, the presence of the Kβ' peak at 2123 eV in the $FePO_4$ spectrum has been shown by DFT calculations to arise from interaction with oxygen in the phosphate bond.[57] In Fig. 7 (bottom), the Kβ emission of GaP is compared to FEFF9 calculations, which uses an *ab initio* multiple-scattering code to calculate the species-specific, local occupied density of states near the Fermi level.[61] After broadening with a Gaussian profile and shifting the energy of the simulated spectrum, there is clear agreement with the measured results.

The ability to select a narrow window of photon energies with our detector results in very low background levels. For the strong P Kα emissions, no appreciable background is observed. For the weaker P Kβ emission, taking the $FePO_4$ emission shown in Fig. 7 as an example, the total, time-integrated background in one energy bin is 22 counts, compared to the 355 counts of the fluorescence line peak intensity at 2138 eV.

In Fig. 8, S Kα measurement results are shown for a ZnS single crystal, using 1300 s measurement time to obtain sufficiently quiet data to also resolve the components of the $K\alpha L^1$ satellite lines[39] on the high-energy tail of the main $K\alpha_{1,2}$ doublet. The Bragg angle for S Kα is much further from back scatter at 59°, and hence a larger bandwidth of 46eV is measured. Due to the small pixels of the camera, small shifts in energy can still be discerned, and high resolution is maintained. Comparing the S Kα spectral shape obtained here with that of Mori *et al.*,[18] good agreement is seen after convolving their spectra with a 0.5 eV FWHM Gaussian profile. Given the reported resolution of 0.44 eV in the prior work, we find an experimental resolution of approximately 0.7 eV for the present instrument in the laboratory environment.

A primary purpose of laboratory-based spectrometers is not to compete with synchrotrons, but instead to enable new directions in analytical chemistry. For a pilot study in such a direction, we use several different preparations of InP QD's to demonstrate the ease with which distributions of oxidation state can be extracted from simple linear superposition fits to reference standards. This is enabled by the significant insensitivity of the spectral shape of the $K\alpha_{1,2}$ doublet to oxidation state – despite overall shifts in energy due to changes in the valence electron population.[18] InP is chosen as a useful representative case not only because of its convenient chemistry in the present context, but also because the surface of InP is readily oxidized, requiring rigorously air-free synthetic procedures for desired applications in solid-state lighting, and biomedical imaging.[62, 63] Hence, a bulk-sensitive quantification of the distribution of P oxidation states is of immediate relevance to characterizing and validating the synthesis process.

In Fig. 9 (middle), InP QDs have been post-synthetically treated with zinc carboxylates leading to a zinc saturated surface environment that has been shown to improve photoluminescence quantum yields.[59] Compared to the as-synthesized InP core in Fig. 9 (top), the amount of oxidized P species increases from 10% to 18% with the addition of zinc. Previous literature has also reported oxidized P levels of InP cores close to 10% by XPS and NMR, but have shown an increase to approximately 35% after shelling with ZnS.[64] In Fig. 9 (bottom), InP magic-sized clusters were prepared with a phosphonate ligand shell, showing the mixed environment of $P^{3-}$ from the inorganic core, and $P^{5+}$ from the organic ligands. While further study, such as cross-comparison to NMR and XPS on the same samples, is needed to fully integrate benchtop XES into this type of analytical approach, the present results strongly support such a campaign. The same approach, if successful, would have even higher impact for analytical chemistry of sulfur-based materials due to the difficulties inherent in applying NMR methods to sulfur.

Finally, the high efficiency and compact size of our DRR spectrometer suggests it can also be easily used at synchrotron and XFEL endstations. At such high-brilliance facilities, there is a wider potential for fundamental application through microprobe studies, time-resolved experiments, or resonant methods capable of more finely interrogating the local electronic structure. Here, we investigate the performance of the miniature DRR spectrometer at the Advanced Light Source (ALS). This comes with the added benefit of a microfocused, monochromatic source, for comparison to the unfocused lab source, and also to determine a well-referenced metric for the instrument's absolute efficiency normalized by unit incident flux, a metric that is difficult to characterize with the laboratory x-ray source. Representative measurements taken at the ALS are included in Figures 6 and 8. The integration time for the GaP spectrum was 7200 s, and for the ZnS was 9600 s. The long measurement times in the synchrotron studies are due to the small flux, i.e., $3 \times 10^9$/s, of the bending-magnet beamline. For both P Kα and S Kα emission, the ALS measurements show slightly better resolution of the $K\alpha_{1,2}$ doublet, but otherwise agree well with the laboratory results. Comparison of the ZnS S Kα signal taken at ALS with results of Mori et al.[18] as above, finds a modest increase in resolution compared to the laboratory, from approximately 0.7 to 0.6 eV.

Compared to the 1200/s for GaP measured in the laboratory, the relatively low flux of the microfocused, bending-magnet source gave a count rate of 7.2/s on the same GaP sample. However, the small-spot illumination of the bending-magnet source (~200μm) improved the spectrometer efficiency by a factor of ~3× because blurring of the point-response function on the sensor was greatly decreased. Normalizing the count rate per unit incident flux ($3\times10^9$/s at 3 keV) we calculate an efficiency of 2.4

counts/$10^9$ incident photons, which compares favorably with that of 0.5 counts/$10^9$ incident photons reported for a prior, high-performing spectrometer at the European Synchrotron Radiation Source.[57] Given that the count rate in a narrow energy band with the microfocused source could be improved by moving the source location farther from the analyzer, i.e., less ‘off-circle’, and given that future implementations of this design may gain a factor of two in count rate by bringing the detector inside the vacuum space, we find good likelihood of long-term impact for this approach at major x-ray facilities, even in the present single-analyzer approach. The extremely compact layout of the optical elements also strongly suggests future multi-analyzer, multi-camera systems to further gain efficiency through multiplexing.

# V. Conclusions

We report an efficient, inexpensive, high-resolution tender x-ray spectrometer that can be used with similar utility in the laboratory with a conventional, low-powered x-ray tube or at synchrotron or x-ray free electron laser endstations. This instrument is the product of the recent commercial availability of small-radius crystal analyzers, our development of a small-pixel energy-resolving x-ray camera, and our choice of a dispersive refocusing Rowland (DRR) geometry that removes much sensitivity to the beamspot size on the sample.

Having particular relevance for future laboratory-based analytical applications of advanced XES, a pilot study was conducted in which the distribution of oxidation states of phosphorous was measured in samples of InP quantum dots having different preparation conditions and consequently different amounts of surface-mediated oxidation. The results suggest high utility for this approach in an analytical chemistry perspective, a venue that will accrue significant benefits especially for sulfur-rich materials, due to the challenges involved in sulfur NMR.

From a scientific and an instrumental perspective, the results presented here suggest many interesting future directions, but we emphasize in closing two instrumental opportunities that we find to be representative. First, while we have emphasized only the demonstrated 2 – 2.5 keV energy range, the performance at higher energy is mostly limited by the decreased quantum efficiency of the present camera above 4 – 5 keV. We have observed XES as high as Fe Kα, but improved camera performance (such as exists with present-day commercial instruments based on CCD sensors) would enable the use of the DRR approach in the lab at higher energies. This will likely require the use of sensors having at least 25 μm pixels, resulting in a reversion to a more typical, ~50 – 100 cm Rowland circle if high resolution is to be maintained. A larger Rowland circle would require a larger instrument, but it would also allow the use of spherical and toroidal analyzers, instead of the present cylindrical optic, to increase collection solid angle. The effective solid angle in a P Kα measurement in the present work is ~ 3 msr, while the solid angle subtended by a conventional 1-m radius spherical analyzer using a 10-cm diameter wafer is 8 msr. Secondly, if one instead retains the current instrument geometry, then the small size of the present instrument suggests a unique scientific advantage: it can be readily integrated into controlled-gas glove box systems to enable new directions in analytical chemistry for air-sensitive materials. This would have wide-ranging use not only for S and P compounds, but also for Tc (whose $L\alpha_{1,2}$ fluorescence lines are in the present energy range) or for M-edge emission of several actinides when using different crystal materials and orientations in the x-ray analyzer.

## Acknowledgements

This work was supported by: the United States Department of Energy, Basic Energy Sciences, under grant DE-SC00008580; the Joint Plasma Physics Program of the National Science Foundation and the Department of Energy under grant DE-SC0016251 (WHM, ORH, GTS); the Heavy Element Chemistry Program at LANL by the Division of Chemical Sciences, Geosciences, and Biosciences, Office of Basic Energy Sciences, U.S. Department of Energy (SAK, ASD); and the U.S. Department of Energy. Los Alamos National Laboratory is operated by Los Alamos National Security, LLC, for the National Nuclear Security Administration of U.S. Department of Energy (contract DE-AC52-06NA25396). Additional support also came from the National Science Foundation under grant number CHE-1552164 (JLS, BMC). This research used resources of the Advanced Light Source, which is a DOE Office of Science User Facility under contract no. DE-AC02-05CH11231.

## References

1. P. Glatzel and U. Bergmann, Coordination Chemistry Reviews **249** (1-2), 65-95 (2005).

2. U. Bergmann and P. Glatzel, Photosynthesis Research **102** (2-3), 255-266 (2009).

3. I. Zaharieva, P. Chernev, G. Berggren, M. Anderlund, S. Styring, H. Dau and M. Haumann, Biochemistry **55** (30), 4197-4211 (2016).

4. J. Kowalska and S. DeBeer, Biochimica et Biophysica Acta (BBA) - Molecular Cell Research **1853** (6), 1406-1415 (2015).

5. Y. Ding, D. Haskel, S. G. Ovchinnikov, Y.-C. Tseng, Y. S. Orlov, J. C. Lang and H.-k. Mao, Phys Rev Lett **100** (4), 045508 (2008).

6. K. M. Davis, M. C. Palenik, L. F. Yan, P. F. Smith, G. T. Seidler, G. C. Dismukes and Y. N. Pushkar, J Phys Chem C **120** (6), 3326-3333 (2016).

7. J. Kern, R. Alonso-Mori, R. Tran, J. Hattne, R. J. Gildea, N. Echols, C. Glockner, J. Hellmich, H. Laksmono, R. G. Sierra, B. Lassalle-Kaiser, S. Koroidov, A. Lampe, G. Y. Han, S. Gul, D. DiFiore, D. Milathianaki, A. R. Fry, A. Miahnahri, D. W. Schafer, M. Messerschmidt, M. M. Seibert, J. E. Koglin, D. Sokaras, T. C. Weng, J. Sellberg, M. J. Latimer, R. W. Grosse-Kunstleve, P. H. Zwart, W. E. White, P. Glatzel, P. D. Adams, M. J. Bogan, G. J. Williams, S. Boutet, J. Messinger, A. Zouni, N. K. Sauter, V. K. Yachandra, U. Bergmann and J. Yano, Science **340** (6131), 491-495 (2013).

8. M. J. Lipp, A. P. Sorini, J. Bradley, B. Maddox, K. T. Moore, H. Cynn, T. P. Devereaux, Y. Xiao, P. Chow and W. J. Evans, Phys Rev Lett **109** (19), 195705 (2012).

9. R. Alonso-Mori, J. Kern, R. J. Gildea, D. Sokaras, T. C. Weng, B. Lassalle-Kaiser, R. Tran, J. Hattne, H. Laksmono, J. Hellmich, C. Glockner, N. Echols, R. G. Sierra, D. W. Schafer, J. Sellberg, C. Kenney, R. Herbst, J. Pines, P. Hart, S. Herrmann, R. W. Grosse-Kunstleve, M. J. Latimer, A. R. Fry, M. M. Messerschmidt, A. Miahnahri, M. M. Seibert, P. H. Zwart, W. E. White, P. D. Adams, M. J. Bogan, S. Boutet, G. J. Williams, A. Zouni, J. Messinger, P. Glatzel, N. K. Sauter, V. K. Yachandra, J. Yano and U. Bergmann, Proceedings of the National Academy of Sciences **109** (47), 19103-19107 (2012).

10. N. Lee, T. Petrenko, U. Bergmann, F. Neese and S. DeBeer, J Am Chem Soc **132** (28), 9715-9727 (2010).

11. H. Gretarsson, A. Lupascu, J. Kim, D. Casa, T. Gog, W. Wu, S. R. Julian, Z. J. Xu, J. S. Wen, G. D. Gu, R. H. Yuan, Z. G. Chen, N. L. Wang, S. Khim, K. H. Kim, M. Ishikado, I. Jarrige, S. Shamoto, J. H. Chu, I. R. Fisher and Y. J. Kim, Phys. Rev. B **84** (10), 100509 (2011).

12. G. Vankó, T. Neisius, G. Molnár, F. Renz, S. Kárpáti, A. Shukla and F. M. F. de Groot, The Journal of Physical Chemistry B **110** (24), 11647-11653 (2006).

13. J. F. Lin, G. Vanko, S. D. Jacobsen, V. Iota, V. V. Struzhkin, V. B. Prakapenka, A. Kuznetsov and C. S. Yoo, Science **317** (5845), 1740-1743 (2007).

14. V. Iota, J. H. P. Klepeis, C. S. Yoo, J. Lang, D. Haskel and G. Srajer, Appl Phys Lett **90** (4), 042505 (2007).

15. M. U. Delgado-Jaime, S. DeBeer and M. Bauer, Chemistry - A European Journal **19** (47), 15888-15897 (2013).

16. C. J. Pollock and S. DeBeer, J Am Chem Soc **133** (14), 5594-5601 (2011).

17. R. A. Mori, E. Paris, G. Giuli, S. G. Eeckhout, M. Kavcic, M. Zitnik, K. Bucar, L. G. M. Pettersson and P. Glatzel, Inorganic Chemistry **49** (14), 6468-6473 (2010).

18. R. A. Mori, E. Paris, G. Giuli, S. G. Eeckhout, M. Kavcic, M. Zitnik, K. Bucar, L. G. M. Pettersson and P. Glatzel, Analytical Chemistry **81** (15), 6516-6525 (2009).

19. Y. Pushkar, X. Long, P. Glatzel, G. W. Brudvig, G. C. Dismukes, T. J. Collins, V. K. Yachandra, J. Yano and U. Bergmann, Angew Chem Int Edit **49** (4), 800-803 (2010).

20. D. R. Mortensen, G. T. Seidler, J. A. Bradley, M. J. Lipp, W. J. Evans, P. Chow, Y. M. Xiao, G. Boman and M. E. Bowden, Review of Scientific Instruments **84** (8), 083908 (2013).

21. A. Leon, A. Fiedler, M. Blum, A. Benkert, F. Meyer, W. L. Yang, M. Bar, F. Scheiba, H. Ehrenberg, L. Weinhardt and C. Heske, J Phys Chem C **121** (10), 5460-5466 (2017).

22. L. Zhang, L. W. Ji, P. A. Glans, Y. G. Zhang, J. F. Zhu and J. H. Guo, Phys Chem Chem Phys **14** (39), 13670-13675 (2012).

23. G. T. Seidler, D. R. Mortensen, A. J. Remesnik, J. I. Pacold, N. A. Ball, N. Barry, M. Styczinski and O. R. Hoidn, Review of Scientific Instruments **85** (11), 113906 (2014).

24. Z. Németh, J. Szlachetko, É. G. Bajnóczi and G. Vankó, Review of Scientific Instruments **87** (10), 103105 (2016).

25. Y. Kayser, W. Błachucki, J. C. Dousse, J. Hoszowska, M. Neff and V. Romano, Review of Scientific Instruments **85** (4), 043101 (2014).

26. L. Anklamm, C. Schlesiger, W. Malzer, D. Grötzsch, M. Neitzel and B. Kanngießer, Review of Scientific Instruments **85** (5), 053110 (2014).

27. I. Mantouvalou, K. Witte, D. Grötzsch, M. Neitzel, S. Günther, J. Baumann, R. Jung, H. Stiel, B. Kanngießer and W. Sandner, Review of Scientific Instruments **86** (3), 035116 (2015).

28. M. Szlachetko, M. Berset, J. C. Dousse, J. Hoszowska and J. Szlachetko, Review of Scientific Instruments **84** (9), 093104 (2013).

29. Y. Gohshi, K. Hori and Y. Hukao, Spectrochim Acta B **B 27** (3), 135-142 (1972).

30. C. Sugiura, Y. Gohshi and I. Suzuki, Phys. Rev. B **10** (2), 338-343 (1974).

31. C. Sugiura, Y. Gohshi and I. Suzuki, Jpn J Appl Phys **11** (6), 911-912 (1972).

32. J. Hoszowska, J. C. Dousse, J. Kern and C. Rhême, Nuclear Instruments and Methods in Physics Research Section A: Accelerators, Spectrometers, Detectors and Associated Equipment **376** (1), 129-138 (1996).

33. M. Kavčič, J. C. Dousse, J. Szlachetko and W. Cao, Nuclear Instruments and Methods in Physics Research Section B: Beam Interactions with Materials and Atoms **260** (2), 642-646 (2007).

34. V. E. Dolgih, V. M. Cherkashenko, E. Z. Kurmaev, D. A. Goganov, E. K. Ovchinnikov and Y. M. Yarmoshienko, Nuclear Instruments and Methods in Physics Research **224** (1-2), 117-119 (1984).

35. Y. M. Yarmoshenko, V. A. Trofimova, L. V. Elokhina, E. Z. Kurmaev, S. Butorin, R. Cloots, M. Ausloos, J. A. Aguiar and N. I. Lobatchevskaya, J Phys Chem Solids **54** (10), 1211-1214 (1993).

36. Y. M. Yarmoshenko, V. A. Trofimova, E. Z. Kurmaev, P. R. Slater and C. Greaves, Physica C **224** (3-4), 317-320 (1994).

37. Y. M. Yarmoshenko, V. A. Trofimova, V. E. Dolgih, M. A. Korotin, E. Z. Kurmaev, J. A. Aguiar, J. M. Ferreira and A. C. Pavao, J Phys-Condens Mat **7** (1), 213-218 (1995).

38. M. Kavcic, A. G. Karydas and C. Zarkadas, X-Ray Spectrom **34** (4), 310-314 (2005).

39. M. Kavčič, A. G. Karydas and C. Zarkadas, Nuclear Instruments and Methods in Physics Research Section B: Beam Interactions with Materials and Atoms **222** (3-4), 601-608 (2004).

40. M. Kavcic, M. Budnar, A. Muhleisen, F. Gasser, M. Zitnik, K. Bucar and R. Bohinc, Review of Scientific Instruments **83** (3), 033113 (2012).

41. T. T. Fister, G. T. Seidler, L. Wharton, A. R. Battle, T. B. Ellis, J. O. Cross, A. T. Macrander, W. T. Elam, T. A. Tyson and Q. Qian, Review of Scientific Instruments **77** (6), 063901 (2006).

42. D. Sokaras, D. Nordlund, T. C. Weng, R. A. Mori, P. Velikov, D. Wenger, A. Garachtchenko, M. George, V. Borzenets, B. Johnson, Q. Qian, T. Rabedeau and U. Bergmann, Review of Scientific Instruments **83** (4), 043112 (2012).

43. D. Sokaras, T. C. Weng, D. Nordlund, R. Alonso-Mori, P. Velikov, D. Wenger, A. Garachtchenko, M. George, V. Borzenets, B. Johnson, T. Rabedeau and U. Bergmann, Review of Scientific Instruments **84** (5), 053102 (2013).

44. R. Verbeni, T. Pylkkänen, S. Huotari, L. Simonelli, G. Vankó, K. Martel, C. Henriquet and G. Monaco, Journal of Synchrotron Radiation **16** (4), 469-476 (2009).

45. O. R. Hoidn and G. T. Seidler, Review of Scientific Instruments **86** (8), 086107 (2015).

46. O. R. Hoidn, W. M. Holden and G. T. Seidler, Review of Scientific Instruments **In preparation** (2017).

47. J. Słowik and A. Zięba, Journal of Applied Crystallography **34** (4), 458-464 (2001).

48. R. W. Cheary and A. Coelho, Journal of Applied Crystallography **27**, 673-681 (1994).

49. S. Brennan, P. L. Cowan, R. D. Deslattes, A. Henins, D. W. Lindle and B. A. Karlin, Review of Scientific Instruments **60** (7), 2243-2246 (1989).

50. E. Welter, P. Machek, G. Drager, U. Bruggmann and M. Froba, Journal of Synchrotron Radiation **12**, 448-454 (2005).

51. S. Huotari, G. Vanko, F. Albergamo, C. Ponchut, H. Graafsma, C. Henriquet, R. Verbeni and G. Monaco, J. Synchrot. Radiat. **12**, 467-472 (2005).

52. S. Huotari, T. Pylkkanen, R. Verbeni, G. Monaco and K. Hamalainen, Nat Mater **10** (7), 489-493 (2011).

53. S. Huotari, F. Albergamo, G. Vanko, R. Verbeni and G. Monaco, Review of Scientific Instruments **77** (5), 053102 (2006).

54. A. C. Hudson, W. C. Stolte, D. W. Lindle and R. Guillemin, Review of Scientific Instruments **78** (5), 053101 (2007).

55. O. R. Hoidn, Ph.D. Thesis, University of Washington, 2017.

56. B. Dickinson, G. T. Seidler, Z. W. Webb, J. A. Bradley, K. P. Nagle, S. M. Heald, R. A. Gordon and I. M. Chou, Review of Scientific Instruments **79** (12), 123112 (2008).

57. M. Petric, R. Bohinc, K. Bucar, M. Zitnik, J. Szlachetko and M. Kavcic, Analytical Chemistry **87** (11), 5632-5639 (2015).

58. D. C. Gary and B. M. Cossairt, Chem Mater **25** (12), 2463-2469 (2013).

59. J. L. Stein, E. A. Mader and B. M. Cossairt, J Phys Chem Lett **7** (7), 1315-1320 (2016).

60. D. C. Gary, M. W. Terban, S. J. L. Billinge and B. M. Cossairt, Chem Mater **27** (4), 1432-1441 (2015).

61. J. J. Rehr, J. J. Kas, F. D. Vila, M. P. Prange and K. Jorissen, Phys Chem Chem Phys **12** (21), 5503-5513 (2010).

62. J. McKittrick and L. E. Shea-Rohwer, J Am Ceram Soc **97** (5), 1327-1352 (2014).

63. P. Mushonga, M. O. Onani, A. M. Madiehe and M. Meyer, J Nanomater, 869284 (2012).

64. H. Virieux, M. Le Troedec, A. Cros-Gagneux, W. S. Ojo, F. Delpech, C. Nayral, H. Martinez and B. Chaudret, J Am Chem Soc **134** (48), 19701-19708 (2012).

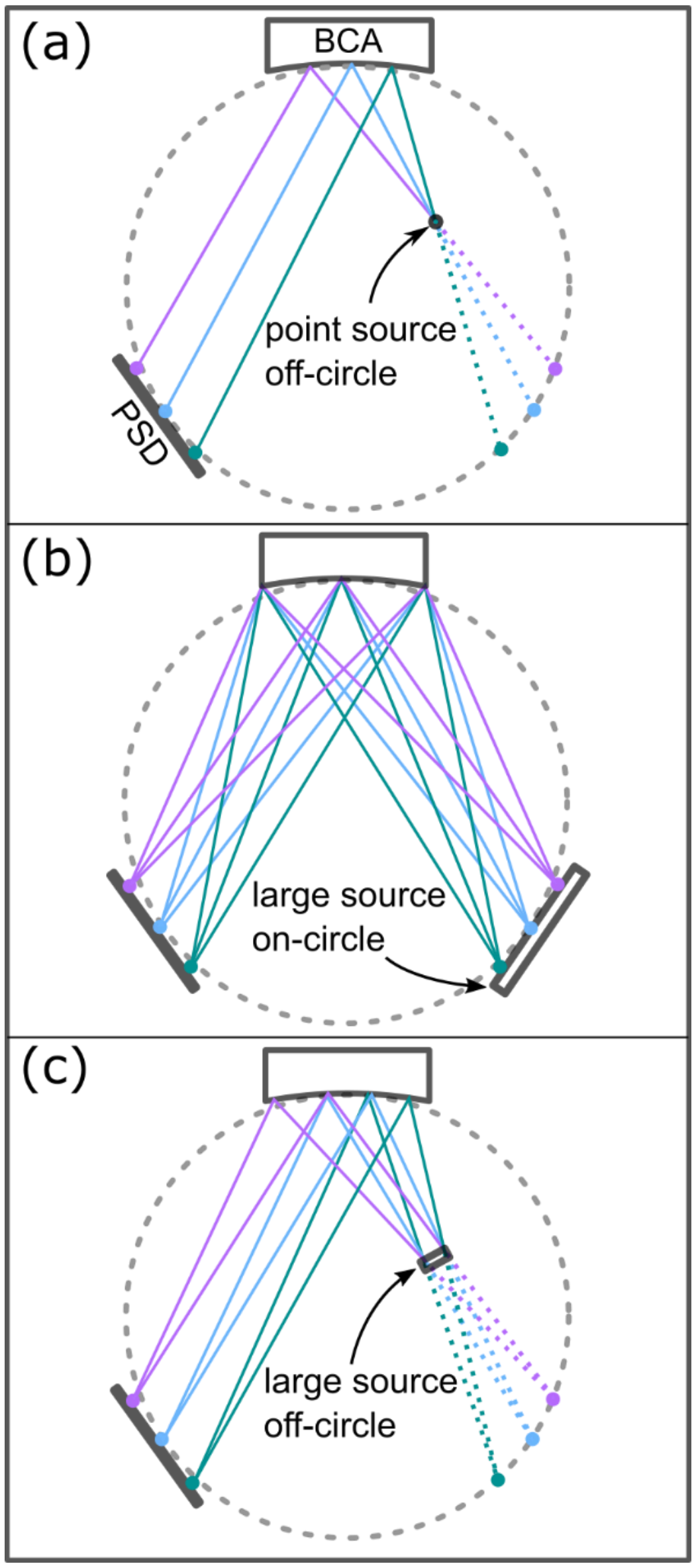

**Figure 1:** Comparison of Rowland circle geometries for different sample illuminations. Each setup has the same basic layout, a source on the lower right arc, a bent crystal analyzer (BCA) at the top of the circle, and a position-sensitive detector (PSD) on the lower left arc. (a) Point source illumination of a sample off of the Rowland circle has 'virtual' rays that can be traced back to intersect the circle. The geometry is dispersive and multiple energies are collected by the PSD. (b) A large source on the Rowland circle has multiple energies that undergo point-to-point focusing and are measured on the PSD. Different regions of the sample contribute at different energies. (c) The dispersive refocusing Rowland (DRR) geometry used in the present work. See the text for discussion.

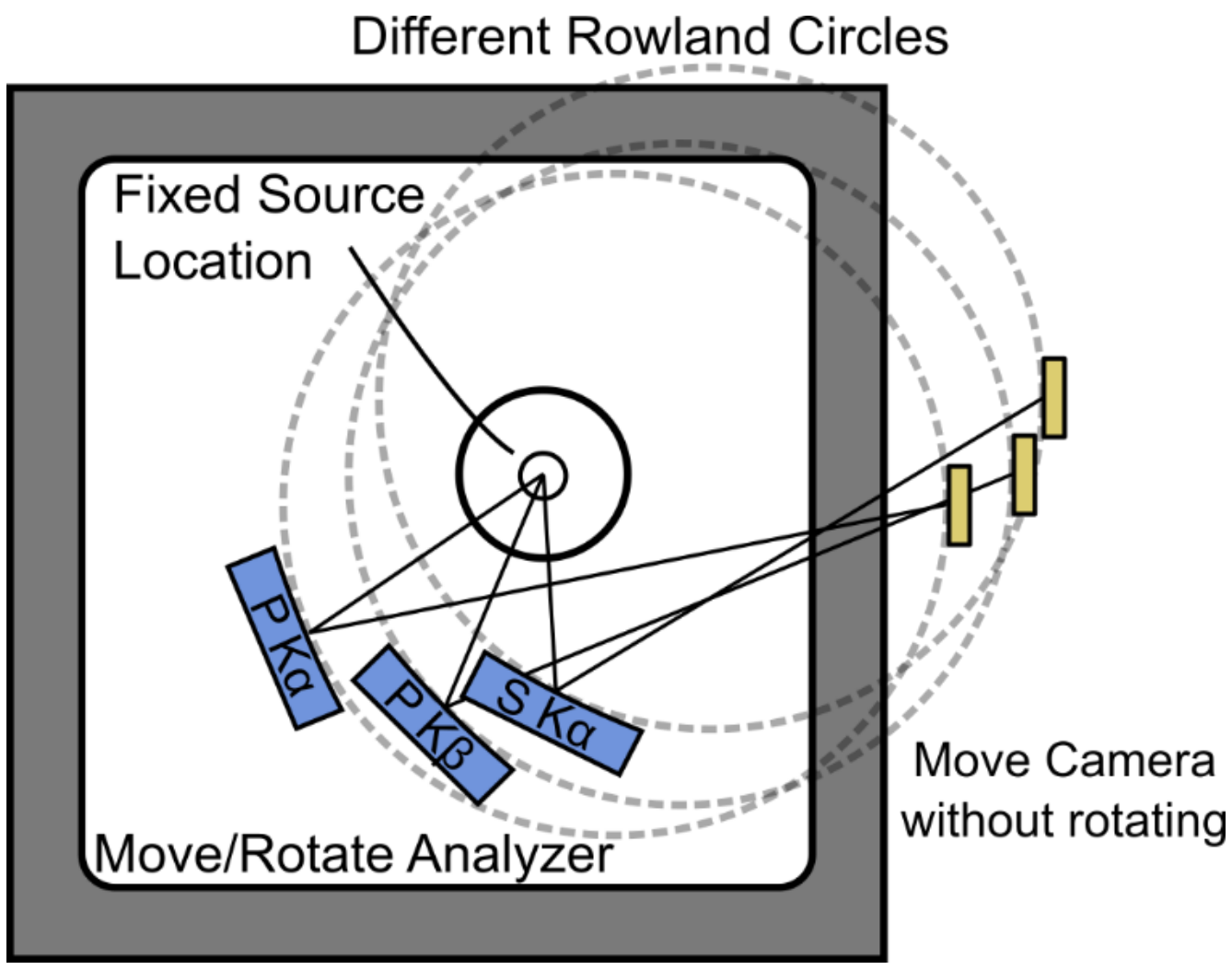

**Figure 2:** Illustration of the implemented DRR design. To fix the source location and maintain camera orientation, the analyzer is moved to different positions yielding different Rowland circles. The sample-analyzer distance is also changed for different Bragg positions to maximize signal from the sample. Maintaining the orientation of the camera reduces the degrees of freedom that need to be optimized to achieve high-resolution.

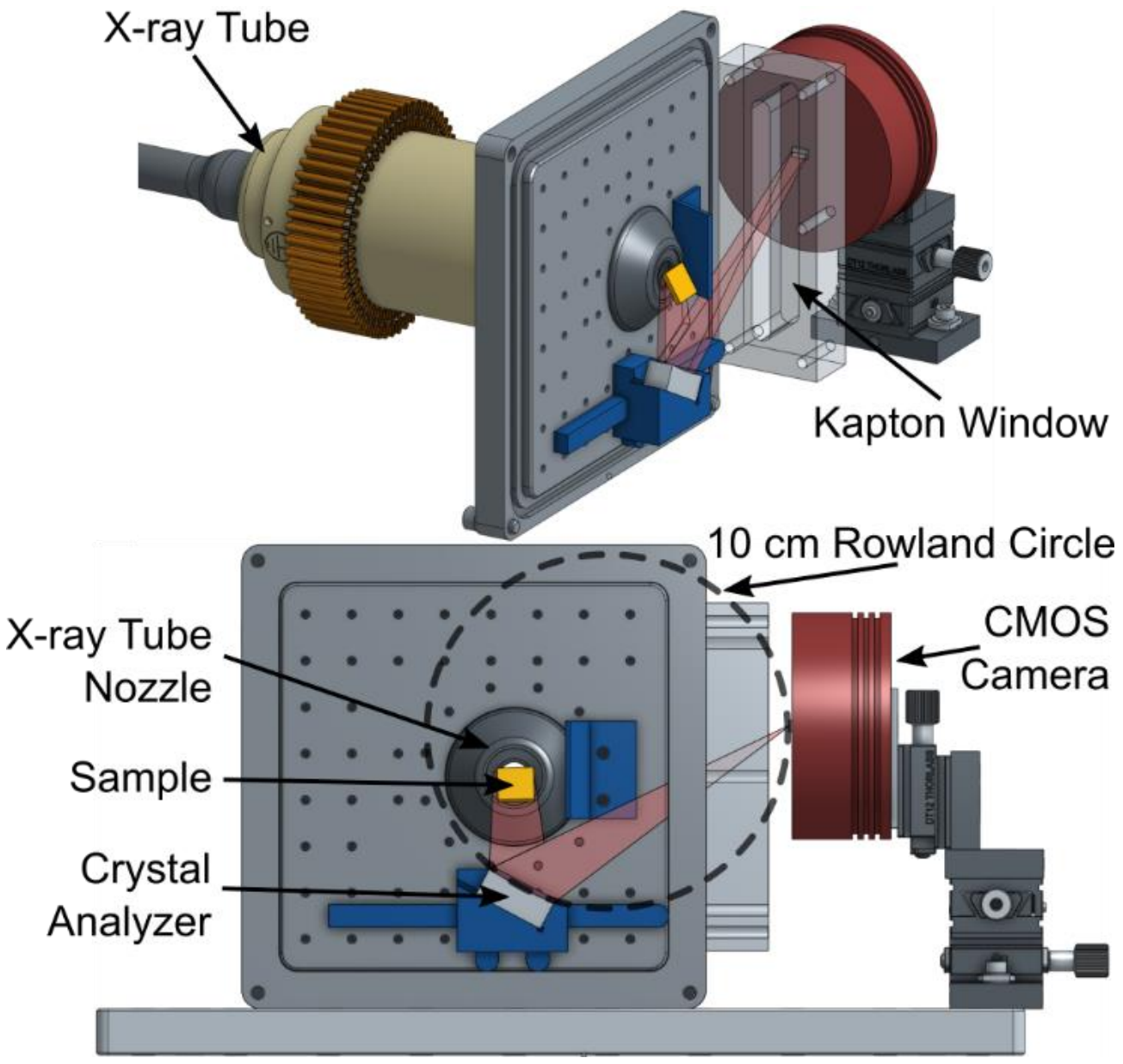

**Figure 3:** Spectrometer CAD renderings illustrating the layout of the components with respect to the Rowland circle. The vacuum chamber has been suppressed for clarity of presentation, see also Fig. 4.

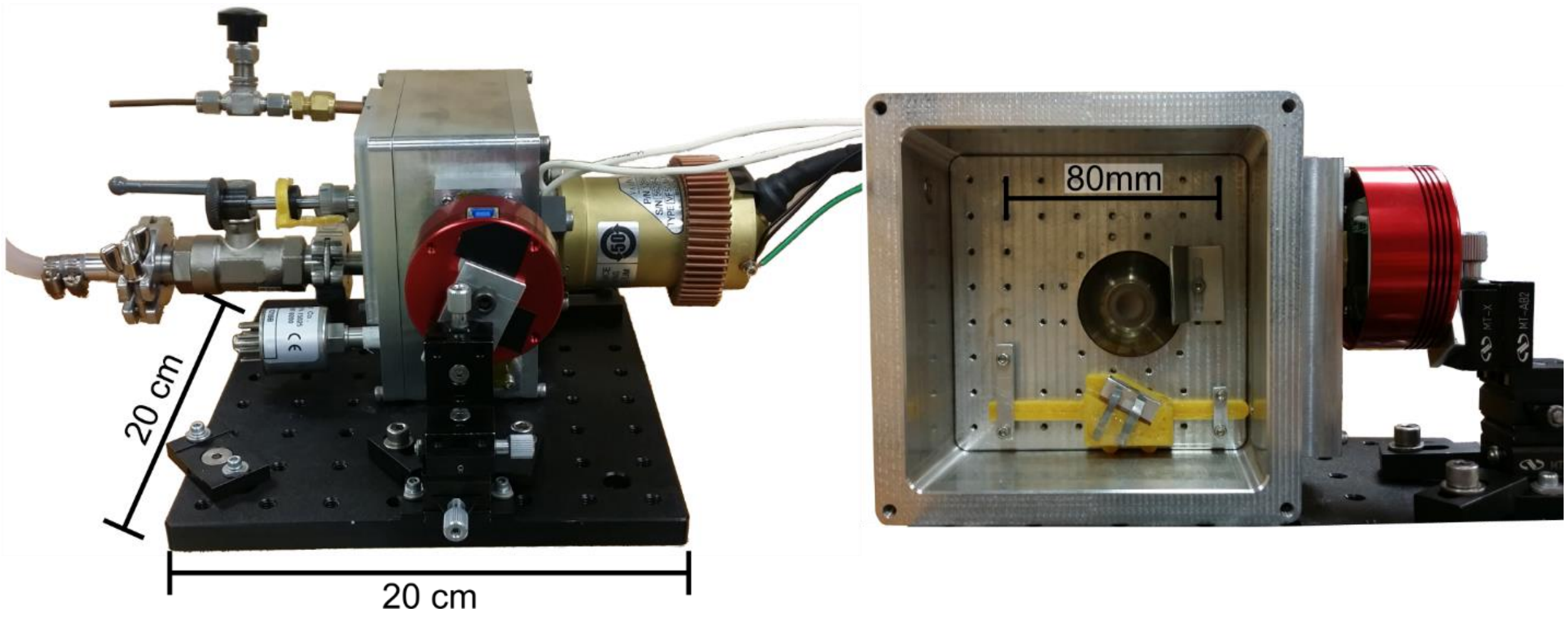

**Figure 4:** Photographs of the spectrometer. Use of a small-radius cylindrically-bent crystal analyzer allows a very small geometry to be utilized. The vacuum chamber houses the nozzle of the x-ray tube, the crystal analyzer, and the sample, which is on a sample turret (not shown) to allow multiple samples to be measured without breaking vacuum or altering the setup.

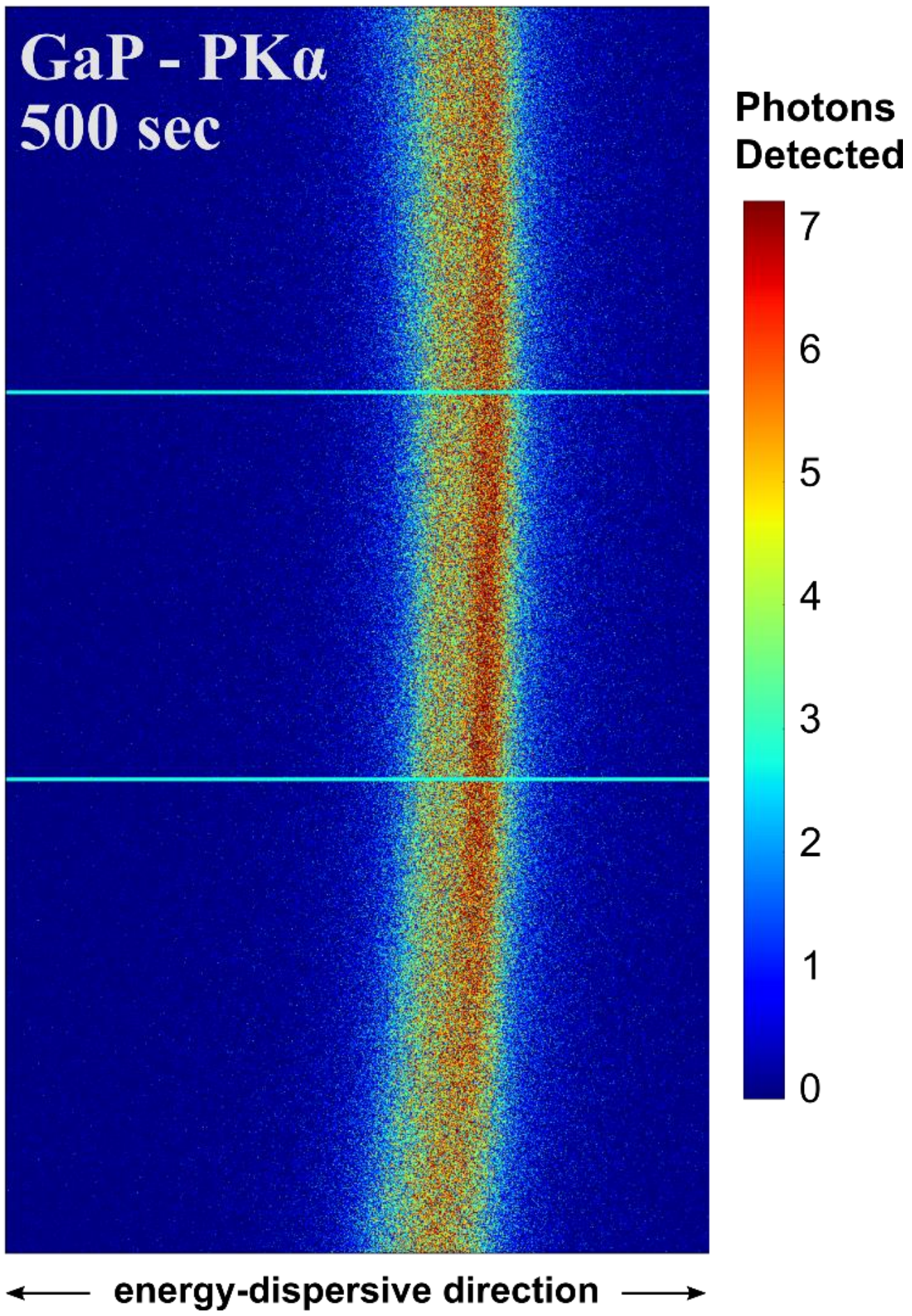

**Figure 5:** Sensor image for data obtained measuring P Kα emission from GaP. The main peak and shoulder of the Kα doublet can be seen. There is an apparent curvature in the data, as well as blurring in the extremes of the curve. For a clean spectrum we process only the cropped, central region indicated by the horizontal lines. The width of the sensor corresponds to a bandwidth of 12 eV at the Bragg angle for P Kα.

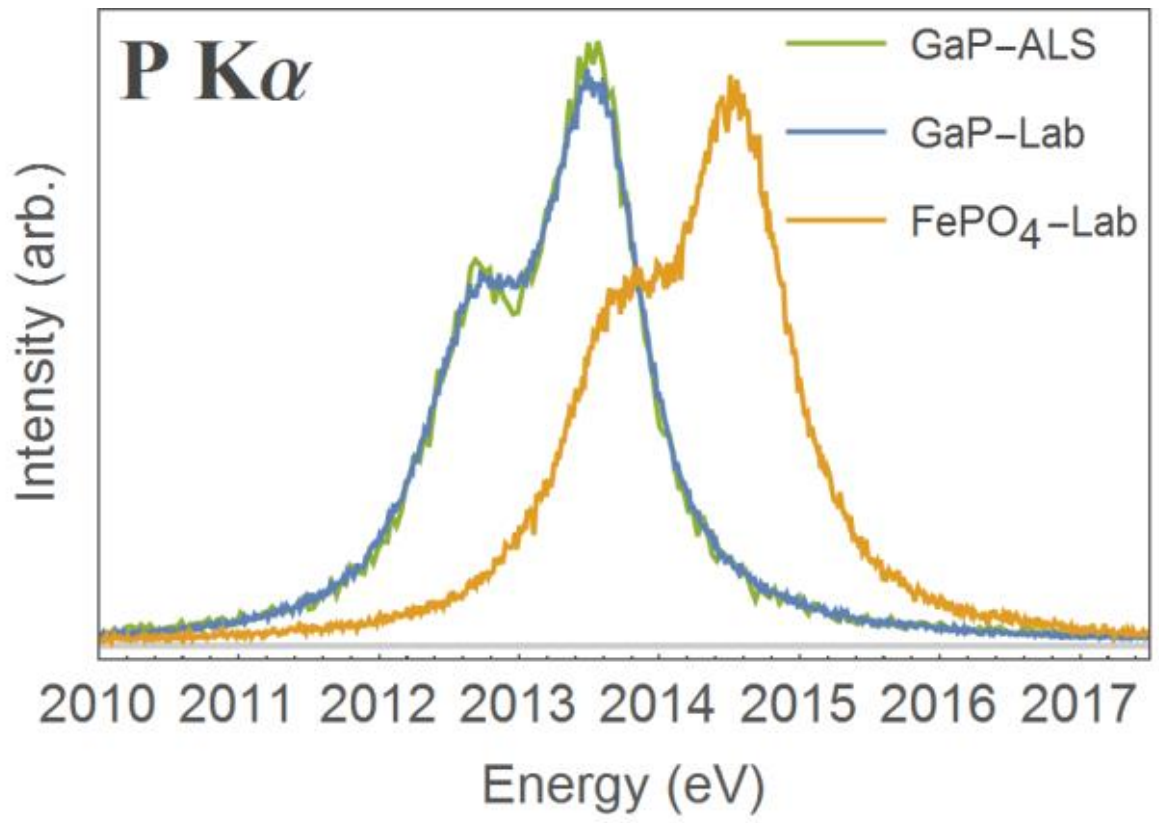

**Figure 6:** Comparison of P Kα emission for samples of GaP and $FePO_4$. The difference in oxidation state ($3^-$ for P in GaP, $5^+$ for P in $FePO_4$) leads to an observed energy shift of the $K\alpha_{1,2}$ doublet by 1.03 eV. For GaP, data is shown for measurements taken in the laboratory and at ALS beamline 10.3.2. The measurement times in the lab for these concentrated samples was 500 s, and the measurement time for GaP at ALS was 7200 s.

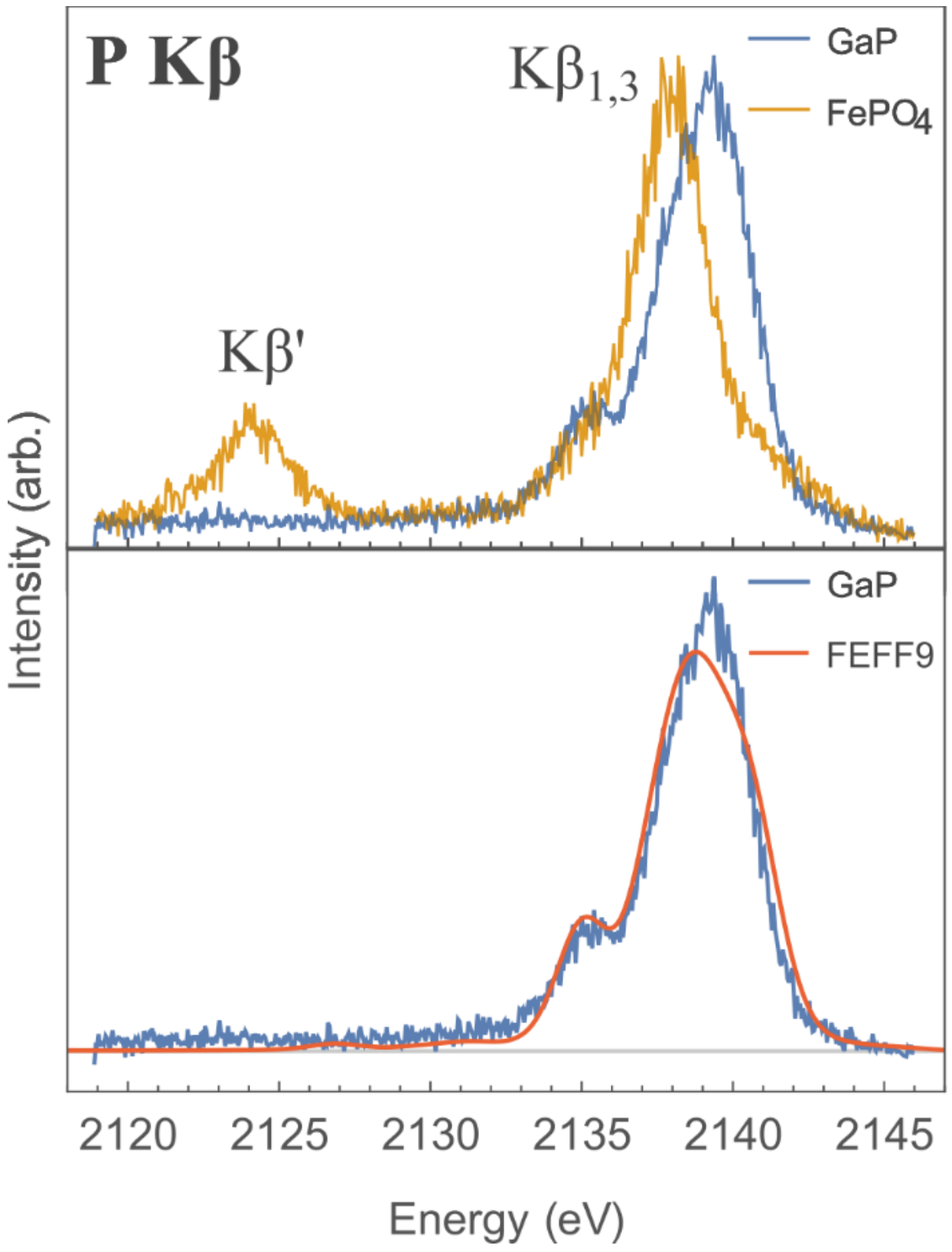

**Figure 7:** Phosphorous Kβ spectrum obtained in the laboratory with a measurement time of 600 seconds. (Top) Comparison of the spectra obtained from samples of GaP and $FePO_{4.}$ The presence and position of the Kβ' peak at 2123 eV in the $FePO_4$ spectrum is a clear indicator of oxygen bonded to the probed P atoms. (Bottom) Comparison of the GaP spectrum to a calculation of the occupied density of states using FEFF9. The output of the FEFF9 calculation was Gaussian broadened and shifted in energy to align with the measured results.

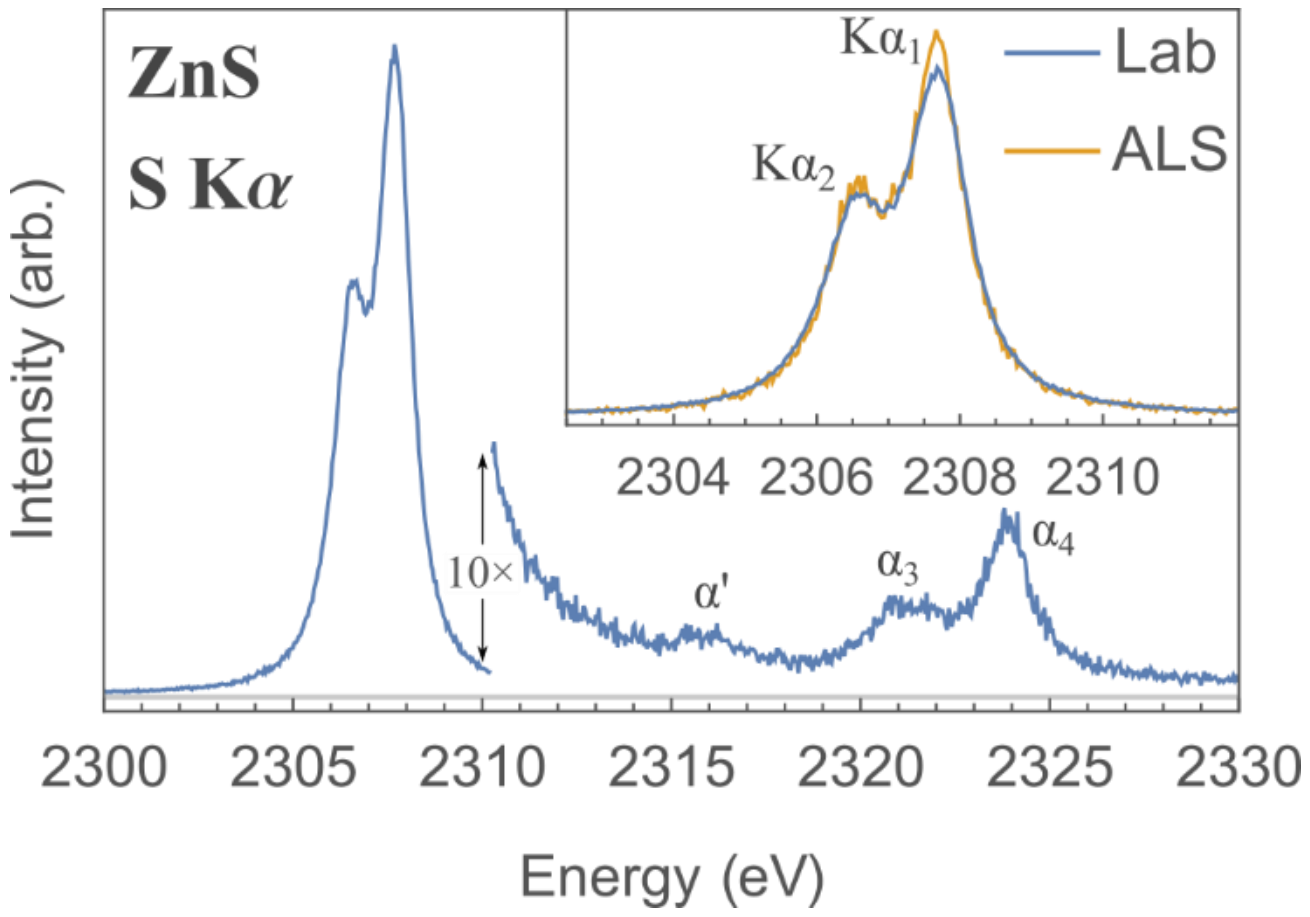

**Figure 8:** Sulfur Kα spectrum of a sample of ZnS, demonstrating the high-resolution capabilities of the spectrometer. Because of the lower Bragg angle, the bandwidth is increased relative to the phosphorous Kα spectrum, and the components of the $K\alpha L^1$ satellite lines on the high-energy tail are clearly resolved. The integration time was 1300 s for the laboratory spectrum, and 9600 s for the spectrum taken at ALS beamline 10.3.2.

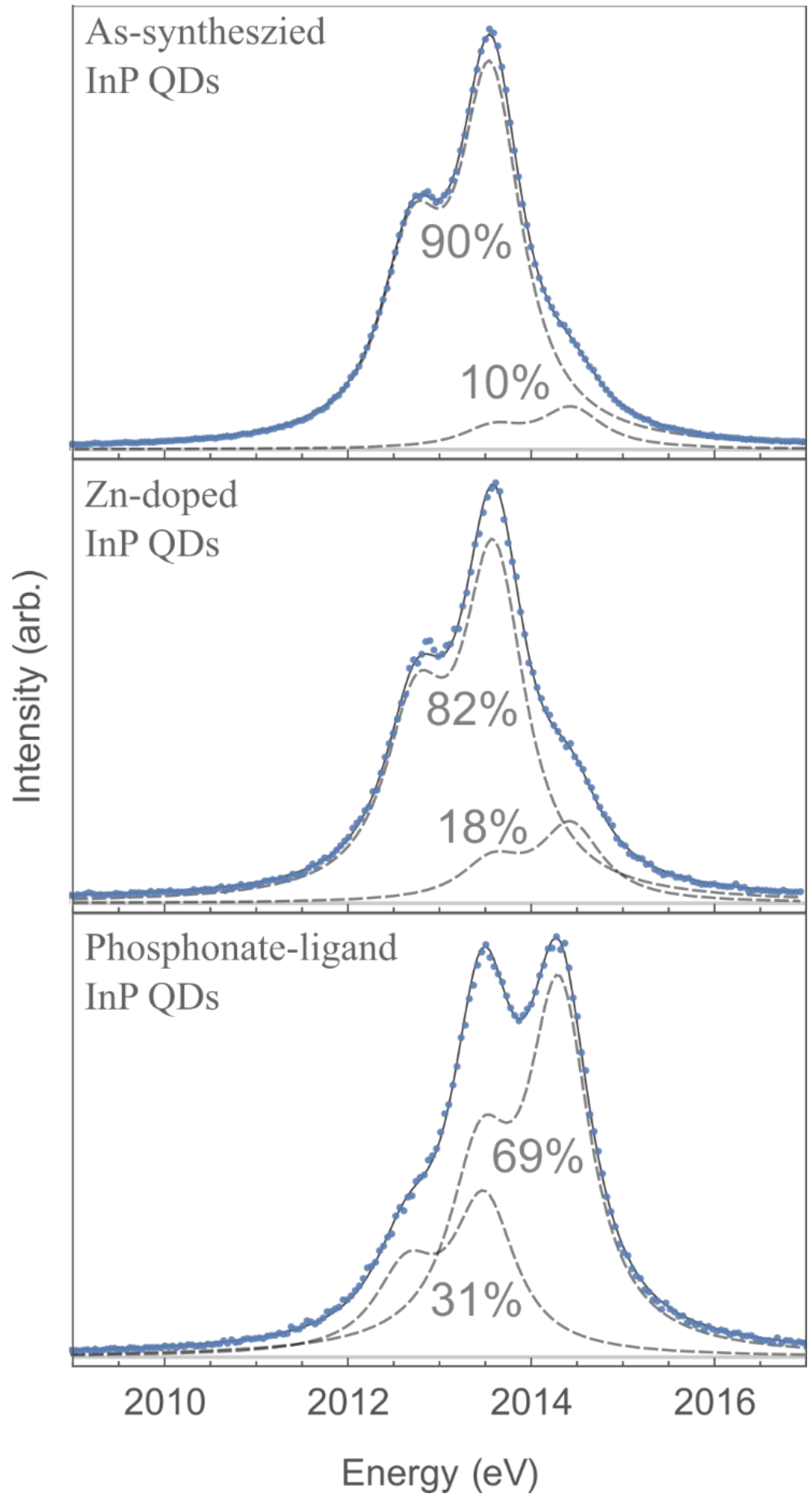

**Figure 9:** Linear combination fits of reference spectra to the P Kα spectra for three different samples of InP nanoparticles showing the relative proportions of oxidation states of phosphorous. The lower energy doublet corresponds to the reduced $P^{3-}$ state of P in InP, while the higher energy doublet represents a highly oxidized state, close in energy to $P^{5+}$ of P in $PO_4^{3-}$. All measurement times were less than 3600 s.